# Blockchain for Academic Integrity: Developing the Blockchain Academic Credential Interoperability Protocol (BACIP)

Author: Juan Alamiro Berrios Moya

AI & Software engineer

Master of Applied Information Technology

VICTORIA UNIVERSITY, AUSTRALIA

https://orcid.org/0009-0007-4713-3504

# ABSTRACT

This research introduces the Blockchain Academic Credential Interoperability Protocol (BACIP), designed to significantly enhance the security, privacy, and interoperability of verifying academic credentials globally, addressing the widespread issue of academic fraud. BACIP integrates dual blockchain architecture, smart contracts, and zero-knowledge proofs to offer a scalable and transparent framework aimed at reducing fraud and improving the mobility and opportunities for students and professionals worldwide.

The research methodology adopts a mixed-methods approach, involving a rigorous review of pertinent literature and systematic integration of advanced technological components. This includes both qualitative and quantitative analyses that underpin the development of a universally compatible system. Preliminary evaluations suggest that BACIP could enhance verification efficiency and bolster security against tampering and unauthorized access.

While the theoretical framework and practical implementations have laid a solid foundation, the protocol's real-world efficacy awaits empirical validation in a production environment. Future research will focus on deploying a prototype, establishing robust validation policies, and defining precise testing parameters. This critical phase is indispensable for a thorough assessment of BACIP's operational robustness and its compliance with international educational standards.

This work contributes significantly to the academic field by proposing a robust model for managing and safeguarding academic credentials, thus laying a strong foundation for further innovation in credential verification using blockchain technology.

# GLOSSARY AND LIST OF ACRONYMS

Acronyms

**AES:** Advanced Encryption Standard

**BACIP:** Blockchain Academic Credential Interoperability Protocol

**EBSI:** European Blockchain Services Infrastructure

**GDPR:** General Data Protection Regulation

**IPFS:** InterPlanetary File System

**JSON-LD:** JavaScript Object Notation for Linked Data

**UUID:** Universally Unique Identifier

**zkEVM:** Zero-Knowledge Ethereum Virtual Machine

Glossary Terms

**Blockchain:** A decentralized ledger that records transactions across multiple computers so that the record cannot be altered retroactively without altering all subsequent blocks.

**Cryptography:** The study of secure communication protocols that prevent third parties or the public from reading private messages.

**Decentralized Storage:** Data storage across a decentralized network to increase data security and availability.

**Digital Certificates:** Blockchain-based digital certificates that verify the authenticity of an academic credential.

**Digital Identity:** Digital identities that allow users to verify and manage their credentials securely.

**Interoperability:** The ability of different systems to work together seamlessly.

**Privacy:** The right of individuals to keep their personal information private and control over their information.

**Security:** Protection of information systems against unauthorized access or modifications.

**Smart Contracts:** Automated contracts that execute under specified conditions within a blockchain.

**Zero-Knowledge Proofs:** Cryptographic methods that prove possession of information without revealing that information.

# TABLE OF CONTENTS

# 1 INTRODUCTION

Academic fraud, particularly through fake degrees and credentials, is a global issue with deep historical roots and various manifestations such as diploma mills and contract cheating. This phenomenon, exacerbated by the high value of educational qualifications and the substantial costs of legitimate education, presents significant challenges in regulation (Jimu, 2012; Eaton & Carmichael, 2023). The ease with which fake degrees can be acquired, thanks to the availability of advanced fraudulent techniques (Børresen et al., 2020), not only devalues authentic academic achievements but also undermines trust in educational systems, affecting professional sectors where expertise is crucial (Attewell & Domina, 2011). Furthermore, these fraudulent practices impose considerable costs on legitimate educational institutions and society, including the loss of tuition revenue and potential damage to the reputation of educational institutions (Brown, 2006; Grolleau, Lakhal & Mzoughi, 2016).

In response, there has been a growing interest in the use of advanced technologies such as blockchain and smart contracts to address diploma forgery. Recent studies indicate a trend toward developing more secure, transparent, and efficient systems for the authentication of academic credentials, proposing blockchain technology as a key tool to ensure the authenticity and integrity of academic records (Michoulis et al., 2022; Ghazaliand & Saleh, 2018; Tang, 2021). Despite these advancements, significant challenges remain such as the need for greater adoption by educational institutions, interoperability between different systems and platforms, and compliance with data privacy and security regulations.

The Blockchain Academic Credential Interoperability Protocol (BACIP) is proposed as a solution to these issues, aiming to facilitate universally recognizable and verifiable academic credential verification, regardless of the specific system each institution uses. This protocol seeks to overcome the barrier of interoperability among various blockchain systems, a notable challenge identified by Capece et al. (2020), promoting broader adoption of this technology in higher education and facilitating the global recognition of digital diplomas.

## 1.1 GENERAL OBJECTIVE
Develop and evaluate the Blockchain Academic Credential Interoperability Protocol (BACIP), designed to enhance interoperability and security in the verification of academic credentials, facilitating their recognition on a global scale.

## 1.2 SPECIFIC OBJECTIVES

### 1.2.1 Define the Core Principles of BACIP
Establish and articulate the key principles of interoperability, security, privacy, transparency, verifiability, scalability, usability, adaptability, evolution, and sustainability that will guide the development of the protocol.

Conduct a comprehensive review of the literature and theoretical frameworks to ensure that BACIP is underpinned by a solid theoretical base relevant to current and future needs in credential verification.

### 1.2.2 Integrate Operational Components of BACIP

Explore and integrate recent innovations in blockchain-based credential verification systems that align with BACIP's principles.

Select and implement technical solutions that enhance the protocol's integrity, efficiency, and accessibility.

### 1.2.3 Standardize Technical Specifications of BACIP

Define and standardize the technical components of BACIP, including architectures, data models, smart contracts, consensus mechanisms, and security protocols based on empirical studies and relevant literature.

Ensure these components effectively address the unique challenges of academic credential verification and are in strict alignment with BACIP's guiding principles.

### 1.2.4 Evaluate Ethical Considerations and Viability of BACIP

Analyze and assess the ethical implications of the protocol, particularly regarding data security and user privacy.

Test the effectiveness and compatibility of the protocol with existing systems, gathering insights that will inform further refinements to ensure that BACIP is not only technically sound but also ethically responsible and practically viable.

## 1.3 RESEARCH QUESTIONS

What technical and operational components are necessary to develop a blockchain protocol (BACIP) that enhances the interoperability and security of academic credential verification?

How can BACIP be evaluated in terms of its effectiveness in improving the security and interoperability of credential verification systems?

What are the potential ethical implications of implementing BACIP, particularly concerning data security and privacy, and how can these be mitigated?

## 1.4 HYPOTHESIS

"The implementation of the Blockchain Academic Credential Interoperability Protocol (BACIP) significantly improves the security and interoperability of academic credential verification across diverse educational systems, contributing to a reduction in academic fraud and an increase in the global mobility of students and professionals."

## 1.5 KEY ACHIEVEMENT AND SIGNIFICANCE

The development and integration of the Blockchain Academic Credential Interoperability Protocol (BACIP) within academic systems represent a significant achievement by providing a decentralized solution that ensures the authenticity and integrity of academic records across various educational platforms. The protocol's ability to facilitate a universal framework for credential verification revolutionizes the traditional methods, significantly reducing the potential for fraud and enhancing the mobility of students and professionals globally. This system's capability to connect different educational and professional entities through a secure and verifiable digital ledger adds significant value to the academic and employment sectors, ensuring that credentials are not only recognized globally but are also resistant to tampering and forgery.

## 1.6   Innovation in the Method

The Blockchain Academic Credential Interoperability Protocol (BACIP) leverages blockchain technology to enhance the verification of academic credentials, emphasizing a well-defined approach based on essential principles such as interoperability, security, privacy, and scalability. A notable innovation within BACIP is its proposal for technical standardization, aimed at establishing a uniform protocol that enhances communication between different educational platforms. This standardization facilitates smoother data exchange and greater compatibility across diverse systems, significantly improving the efficacy of credential verification processes. By integrating a dual blockchain architecture, BACIP not only ensures the secure and private management of credentials but also supports transparent verification processes across public and private networks. Moreover, the use of smart contracts and zero-knowledge proofs further secures and streamlines these interactions, making BACIP a robust framework for global educational institutions.

## 1.7   Potential Improvement to the Method

Recognizing the need for empirical validation, it is proposed to implement a prototype of the BACIP to assess its effectiveness under real conditions. This step is crucial to identify and address operational and technological challenges that may not be evident in the theoretical phase. The evaluation of the prototype will provide valuable data to refine the protocol, enhance interoperability and usability, and ensure its compliance with current and future privacy and security regulations. This practical approach will ensure that the protocol can be effectively adapted to the changing needs of educational institutions and regulatory environments.

## 2 PRELIMINARY LITERATURE REVIEW

Academic fraud in higher education, particularly through fake degrees and credentials, is a global issue with deep historical roots, taking various forms such as diploma mills and contract cheating, as discussed by Eaton & Carmichael (2023). This phenomenon is fuelled by the high value of educational qualifications and the substantial costs of legitimate education, facing significant challenges in regulating this issue (Jimu, 2012). It is revealed that acquiring fake degrees is disturbingly easy, due to the widespread availability of diploma mills and advanced fraudulent techniques (Børresen et al., 2020), devaluing authentic academic achievements and posing risks in professional sectors where expertise is crucial. The prevalence of such fraudulent practices aligns with sociological theories of deviance, reflecting broader socio-economic pressures and undermining societal trust in educational systems (Attewell & Domina, 2011), highlighting a critical need for effective detection and prevention measures to uphold the standards and credibility of higher education worldwide.

Further, fake diplomas constitute a complex and far-reaching problem that affects multiple aspects of society and the economy globally (Grolleau, Lakhal & Mzoughi, 2016). These inauthentic credentials not only represent a significant source of income for the organizations producing them but also generate considerable costs for legitimate educational institutions and society at large, including the loss of tuition revenue for real universities, costs associated with protecting intellectual property rights, and potential damage to the reputation of higher education institutions (Brown, 2006). The proliferation of fake degrees erodes trust in the educational system, leading to significant challenges for employers and other stakeholders in discerning between legitimate and false qualifications (Eaton & Carmichael, 2023), leading to a general devaluation of academic degrees and diminishing their value in the job market and society (Firmo, 2021). In the workplace, fake diplomas enable individuals without the proper qualifications to gain employment or advance in their careers, raising competency and efficacy issues, and highlighting the need for more effective policies and technologies to combat this type of fraud (Brown, 2006; Firmo, 2021).

Moreover, the analysis of literature on the verification of academic diplomas using advanced technologies, particularly blockchain technology and smart contracts, reveals an increasing and diversified focus on finding solutions to the problem of diploma forgery (Ghazaliand & Saleh, 2018; Tang, 2021). Through various studies, a trend towards the development of more secure, transparent, and efficient systems for the authentication of academic credentials is highlighted, with blockchain technology proposed as a key tool for ensuring the authenticity and integrity of academic records (Michoulis et al., 2022). Blockchain-based systems facilitate the creation of a decentralized and secure registry of diplomas and academic certificates, making their alteration or forgery more difficult. The transparency of the blockchain allows different parties, such as educational institutions, employers, and the students themselves, to access and verify the authenticity of diplomas efficiently. Smart contracts, operating on platforms like Ethereum, enable the automation of the process of issuing, storing, and verifying diplomas, making the process more efficient and reducing manual intervention (Pathak et al., 2022). While these research efforts present an encouraging outlook for the use of blockchain technology and smart contracts in combating diploma forgery, challenges remain, such as the need for greater adoption by educational institutions, interoperability between different systems and platforms, and compliance with data privacy and security regulations (Ghazaliand & Saleh, 2018; Tang, 2021), paving the way for future research and developments focused on the implementation of existing technology and considering key points that require improvement. Institutions like MIT and the University of Rome "Tor Vergata" have spearheaded the implementation of blockchain technology for issuing and verifying diplomas, marking significant strides in the digitalization of academic credentials. Despite these advancements, interoperability among various blockchain systems stands out as a notable challenge (Capece et al., 2020). This challenge focuses on the ability of different blockchain systems to interact seamlessly, enabling universal credential verification regardless of the specific system each institution uses. Overcoming this hurdle is crucial for expanding blockchain technology's potential in higher education, encouraging broader adoption, and facilitating the global recognition of digital diplomas.

## 3 THEORETICAL FRAMEWORK

### 3.1 INFORMATION AND COMMUNICATION THEORIES

In the realm of distributed ledger technologies, information and communication theories play a crucial role in conceptualizing secure and efficient systems. Cryptography theory, for instance, relies on mathematical principles to ensure the confidentiality, integrity, authenticity, and non-repudiation of digital information (Diffie & Hellman, 1976). This theory is foundational to the security of blockchain technologies, allowing transactions to be secure and publicly verifiable without the need for a central authority (Nakamoto, 2008).

On the other hand, game theory provides a framework for analyzing strategic interactions among participants in a decentralized system. This theory is applied to ensure that, despite the selfish actions of individuals, the blockchain system maintains its integrity and security (Buterin, 2013). For example, the Proof of Work consensus mechanism can be understood through game theory as an effective compromise between system security and resource consumption (Narayanan et al., 2016).

Furthermore, Shannon's information theory (1948) offers a framework for understanding how data transmission efficiency can be measured and maximized in blockchain systems. Redundancy and error correction are key concepts derived from this theory, essential for the robustness of blockchains in transmitting and storing information (Shannon, 1948).

### 3.2 FUNDAMENTAL PRINCIPLES OF BLOCKCHAIN TECHNOLOGY

Blockchain technology, grounded in a set of interdisciplinary theoretical principles, has revolutionized the way we understand and manage digital information in a decentralized manner. Below are some of the core principles underlying this technology:

#### 3.2.1 Decentralization and Data Autonomy

Unlike traditional centralized systems, blockchain is characterized by its decentralized structure, where information is not stored at a single point but is distributed among all participants in the network (Tapscott & Tapscott, 2016). This feature ensures that no entity or individual has absolute control over the system, promoting greater transparency and resistance to censorship and cyber attacks.

#### 3.2.2 Immutability and Traceability

Blockchain ensures the immutability of records; once information has been added to the chain, it cannot be altered or deleted. This is made possible by cryptographic algorithms that link each block to its predecessor through a unique cryptographic hash, thus creating a clear and verifiable history of all transactions (Narayanan et al., 2016). This property is essential for trust in systems where the veracity of the transaction history is critical, such as in the financial sector.

#### 3.2.3 Distributed Consensus

The consensus mechanism is key to the operation of blockchains, allowing nodes in the network to agree on the current state of the chain in a democratic manner and without the need for a central authority. There are various consensus algorithms, such as the Proof of Work (PoW), Proof of Stake (PoS), and other hybrid models, each with its advantages in terms of energy efficiency, security, and scalability (Buterin, 2013; King & Nadal, 2012).

### 3.2.4 Transparency and Privacy

Although all transactions on a blockchain are public and verifiable by any network participant, the technology also offers tools to ensure user privacy. Through techniques such as pseudonymous addresses and public key cryptography, users can conduct transactions without revealing their real identities, thus balancing transparency with the need for privacy (Nakamoto, 2008).

### 3.2.5 Interoperability and Open Standards

The evolution of blockchain technology has led to the development of open standards and interoperability protocols that facilitate communication and transaction between different blockchains and external systems. This is crucial for the creation of more complex blockchain networks and for their integration into a wide range of applications, from finance and logistics to electronic voting and beyond (Buterin, 2015).

These principles not only define the technical operation of blockchain technology but also reflect a paradigm shift in the management of digital information, promoting a more secure, transparent, and equitable environment.

## 3.3 SECURITY AND PRIVACY IN DISTRIBUTED SYSTEMS

Security and privacy are foundational elements in the design and implementation of distributed systems, particularly in critical applications such as the verification of academic credentials via blockchain technologies. The integrity of the blockchain system and the confidentiality of user data are essential to maintain trust in the credential verification and validation processes.

### 3.3.1 Security in Blockchain

Security mechanisms in blockchain, specifically designed to ensure interoperability, play a crucial role in maintaining a secure ecosystem for the exchange of academic credentials. Among these techniques, Zero-Knowledge Proofs (ZKPs) stand out for their ability to verify the authenticity of a claim without revealing the underlying information (Goldwasser, Micali, & Rackoff, 1989). This property is particularly valuable in the academic context, where credential verification must balance transparency with privacy. For example, an institution could verify that an individual possesses a relevant degree without needing direct access to the student's academic records, thereby preserving their privacy.

Moreover, the immutability and transparency inherent to blockchain technology provide a robust mechanism to prevent the alteration and forgery of academic records, ensuring that only verified and legitimate credentials are recognized throughout the system. This decentralized approach not only enhances data security but also facilitates interoperability among various institutions and blockchain platforms (Narayanan et al., 2016).

### 3.3.2 Privacy and Data Protection

In the exchange of academic credentials, the protection of personal data is paramount. The European Union's General Data Protection Regulation (GDPR) sets stringent guidelines for the handling of personal data, including the right to be forgotten, which presents unique challenges for the implementation of blockchain technologies characterized by their immutable nature (Voigt & Von dem Bussche, 2017). However, blockchain solutions can be designed to be GDPR-compliant, for example, by utilizing data structures that allow for the revocation of access without altering the blockchain's integrity. This can be achieved through data encryption and the use of links to personal data stored off-chain, rather than on the blockchain itself.

Therefore, the implementation of blockchain technologies in the academic realm must carefully consider personal data management strategies to align technical capabilities with legal and ethical requirements. Adopting a privacy-by-design approach from the onset of developing blockchain solutions for academic credential verification is crucial to ensuring compliance with data protection regulations and safeguarding the rights of students and professionals (Schwerin, 2018).

## 3.4 PROTOCOL THEORY

Communication protocol theory encompasses the set of rules and norms that govern the exchange of information between different systems and networks. These protocols ensure that communication is efficient, secure, and reliable, essential elements in any information technology system (Tanenbaum & Wetherall, 2011). In the context of blockchain, protocols not only facilitate data transmission but also incorporate advanced security and privacy mechanisms, crucial for the protection of transactions and sensitive information (Narayanan et al., 2016).

### 3.4.1 Types of Protocols and Their Relevance to BACIP

Within the blockchain realm, there are various types of protocols, each with specific characteristics designed to solve different communication and security problems. For example, consensus protocols, like PoW and PoS, are crucial for validating transactions without a centralized authority, while cryptographic protocols, such as digital signatures and ZKPs, ensure data security and privacy (Buterin, 2014; Goldwasser, Micali, & Rackoff, 1989). The development of BACIP involves selecting and adapting these protocols to meet the specific needs of academic credential verification and interoperability.

### 3.4.2 Interoperability in Protocols

Interoperability among different blockchain systems is a significant challenge that BACIP aims to address. Adopting open standards and creating compatible protocols are essential for enabling the fluid exchange of academic credentials globally (Swan, 2015). This need underscores the importance of designing BACIP with a focus on compatibility and adopting standard practices in protocol development.

### 3.4.3 Challenges and Considerations in Protocol Development

Designing specific protocols for the interoperability of academic credentials faces several challenges, including data security, user privacy, and long-term sustainability. It is crucial for BACIP to address these challenges by implementing advanced technical solutions and adhering to regulations like the GDPR for data protection (Voigt & Von dem Bussche, 2017).

# 4 METHODOLOGY

## 4.1 METHODOLOGICAL FRAMEWORK

In advancing the development and evaluation of the Blockchain Academic Credential Interoperability Protocol (BACIP), this study adopts a mixed-methodology framework, structured into four distinct phases. This approach is meticulously designed to address specific technological requirements and maintain rigorous technical specifications. Simultaneously, it aims to comprehensively assess the protocol's impact, effectiveness, and security in real-world scenarios. The primary focus is on establishing a robust and flexible foundation for BACIP, laying the groundwork for potential future advancements. This foundation is intended to benefit this project and potentially others within the academic and development communities.

### 4.1.1 Phase I: Articulation of BACIP's Core Principles

We begin by defining the fundamental principles that anchor BACIP. This foundational phase is pivotal, as it outlines the principles of Interoperability, Security, Privacy, Transparency and Verifiability, Scalability, Usability, Adaptability and Evolution, and Sustainability. These principles are derived from a comprehensive review of literature and theoretical frameworks, ensuring that BACIP is underpinned by a solid theoretical base. This base not only addresses the existing challenges in credential verification but is also forward-looking, anticipating future developments. We highlight our commitment to grounding BACIP in thorough research, setting the stage for addressing the complexities of today's academic and technological landscapes with an effective and theoretically backed approach.

### 4.1.2 Phase II: Integration of Operational Components

With the foundational principles established, we turn our focus to defining BACIP's operational components. This phase involves an in-depth exploration of recent innovations in blockchain-based credential verification systems, aiming to incorporate components that align with BACIP's principles and address contemporary challenges. Through a rigorous selection process, we aim to integrate solutions that enhance the protocol's integrity, efficiency, and accessibility, thereby ensuring BACIP remains at the forefront of technological advancements in credential verification.

### 4.1.3 Phase III: Standardization and Technical Specifications

In Phase III, the focus narrows to the technical components of BACIP, aiming to establish and standardize these components based on empirical studies and literature that echo BACIP's core principles. This meticulous process involves evaluating and selecting architectures, data models, smart contracts, consensus mechanisms, and security protocols. Our goal is to incorporate components that not only effectively solve the unique challenges of academic credential verification but also are in strict alignment with BACIP's guiding principles, facilitating a protocol that is both innovative and robust.

### 4.1.4 Phase IV: Ethical Considerations and Viability

The final phase extends beyond technicalities to consider BACIP's ethical implications, particularly regarding data security and user privacy. Evaluating the protocol's effectiveness and compatibility with existing systems. Insights gained from this phase will inform further refinements, ensuring that BACIP is not just technically sound but also ethically responsible and practically viable.

### 4.1.5 Schedule

| Week | Activities |
|---|---|
| 1 | Definition of research objectives and scope. |
| 2 | Start of literature review on academic fraud and verification technologies. |
| 3 | Detailed analysis of existing literature and development of a theoretical framework focused on blockchain and smart contracts. |
| 4 | Continuation of literature analysis; identification of existing studies that justify specific methodological and technical decisions. |
| 5 | Presentation of the first progress report to the supervisor, including results from the initial rounds of literature review. |
| 6 | Reevaluation of literature to identify gaps and redefine the thesis proposal, designing a new methodological plan. |
| 7 | Formulation of BACIP principles based on literature and theoretical framework. |
| 8 | Review of new literature focused on current academic accreditation solutions based on blockchain to detect potential operational components of BACIP. |
| 9 | Research for the justification and definition of BACIP's technical specifications. |
| 10 | Evaluation of ethical considerations and viability |
| 11 | Final presentation and defense in front of supervisors, including results and future recommendations. |

Through this structured methodological approach, BACIP is designed to be a pioneering protocol that meticulously addresses the multifaceted nature of academic credential verification. Grounding each development phase in BACIP's core principles and leveraging a wide range of research insights, the protocol is poised to establish a new benchmark for credential verification systems. The methodology framework outlined in **Figure 1** illustrates the sequential phases starting with the articulation of core principles and culminating in a complete methodology, ensuring the robust development and future viability of BACIP.

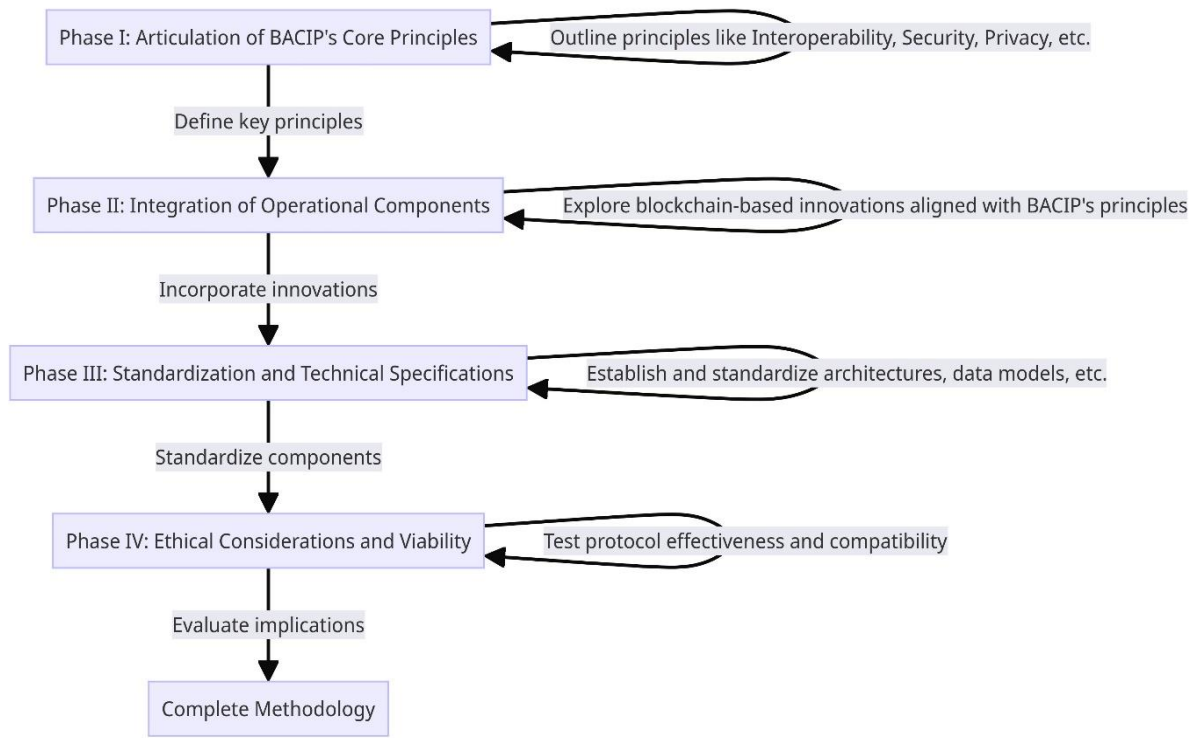

**Figure 1: Methodology Framework for BACIP Development.**

Figure 1 presents the structured methodology framework for the development of the Blockchain Academic Credential Interoperability Protocol (BACIP). It delineates the sequential phases starting with the articulation of core principles and culminating in a complete methodology. Each phase outlines specific objectives, such as defining principles, integrating operational components, standardizing technical specifications, and considering ethical implications, all of which contribute to the robust development and future viability of BACIP.

## 4.2 BACIP PRINCIPLES

The development of the Blockchain Academic Credential Interoperability Protocol (BACIP) is founded on a set of core principles designed to ensure its effectiveness, security, and widespread adoption. These principles guide every stage of the protocol's design, implementation, and evaluation, ensuring that BACIP adequately meets the needs of educational institutions, students, and the employment market. Below are the principles that constitute the backbone of BACIP:

### 4.2.1 Interoperability

BACIP commits to facilitating seamless interoperability among diverse blockchain platforms and academic information systems. This involves developing open standards and employing technologies that allow the fluid exchange of academic credentials across different blockchains and technological environments, thus promoting a globally connected educational network.

### 4.2.2 Security
Security is paramount in BACIP. The protocol utilizes advanced cryptographic techniques to protect the integrity and confidentiality of academic credentials, ensuring that only authorized parties can access them. This principle extends to safeguarding against unauthorized alterations and malicious attacks, providing a trustworthy foundation for all transactions.

### 4.2.3 Privacy
BACIP values and protects the privacy of students' and institutions' data. Mechanisms are implemented that allow for access control to credentials, ensuring that only individuals or entities with permission can view or verify academic information. Additionally, technologies such as Zero-Knowledge Proofs are explored to facilitate credential verification without revealing sensitive information.

### 4.2.4 Transparency and Verifiability
The protocol promotes transparency and the easy verification of the authenticity of academic credentials. Through the use of blockchain, BACIP provides an immutable and public (or private but accessible to authorized parties) record that facilitates the validation of the provenance and validity of academic credentials.

### 4.2.5 Scalability
BACIP is designed to scale efficiently as new educational institutions and users join the system. Special attention is given to the selection of blockchain technologies and the design of smart contracts to handle an increasing volume of transactions without compromising performance.

### 4.2.6 Usability
Ease of use is a central pillar of BACIP, aiming to ensure that the protocol is accessible to all users, regardless of their technical expertise. This includes providing intuitive interfaces through the API and producing detailed documentation and support resources.

### 4.2.7 Adaptability and Evolution
BACIP is a dynamic protocol, capable of adapting to technological, regulatory, and market changes. A proactive approach to innovation and continuous improvement is maintained, allowing the protocol to evolve to meet the emerging needs of the educational and technological ecosystem.

### 4.2.8 Sustainability
Considerations of environmental and economic sustainability guide the development of BACIP, with a focus on optimizing resources and minimizing the operational and environmental costs associated with blockchain transactions.

## 4.3 INTERCONNECTED PRINCIPLES OF BACIP
The foundational principles of the Blockchain Academic Credential Interoperability Protocol (BACIP) stand not only independently but also interweave to form a comprehensive structure ensuring the efficiency and efficacy of the protocol. The synergy and dynamic relationships between these fundamental principles are explored below:

Interoperability, as the cornerstone of BACIP, drives cohesion and compatibility across diverse blockchain platforms and academic information systems. Achieving such interoperability relies on a solid foundation of Security, ensuring communication between systems is immune to unauthorized alterations and access. Privacy, inherently linked to security, is reinforced through mechanisms controlling who can access and verify credentials, thus safeguarding users' personal information.

Transparency and Verifiability naturally flow from the union of security and privacy, enabling credentials to be confidently and effortlessly authenticated. This principle is key in maintaining a clear and trustworthy record of credentials that all stakeholders can validate, thus strengthening the integrity of the entire educational ecosystem.

Scalability is paramount for BACIP's growth and expansion, feeding directly off the principle of security. As more institutions and users adopt the protocol, the infrastructure must handle an increasing volume of data without compromising speed or stability, ensuring that the system is sustainable in the long term.

Usability emerges as a core concern for BACIP, as a complex system is useless if its users cannot efficiently navigate it. Therefore, meticulous attention is paid to crafting interfaces and documentation that make the protocol accessible and understandable for everyone, which in turn fosters greater adoption and user satisfaction.

The principle of Adaptability and Evolution acknowledges that technology and educational needs are constantly evolving, so the protocol is built with the necessary flexibility to adapt and evolve. This principle ensures that BACIP not only meets current needs but also anticipates and adapts to future changes.

Lastly, Sustainability is the culmination of careful consideration of how BACIP operates within the broader context of the environment and economy. By optimizing resources and minimizing costs, it is ensured that the protocol is not only viable in the present but continues to be relevant and effective for future generations.

The interconnections between the core principles of BACIP are depicted in **Figure 2**, which shows how each individual principle contributes to the collective vision of the protocol. This illustration reinforces the robustness and reliability of BACIP and its capacity to adapt to the evolving demands of the educational and technological landscape.

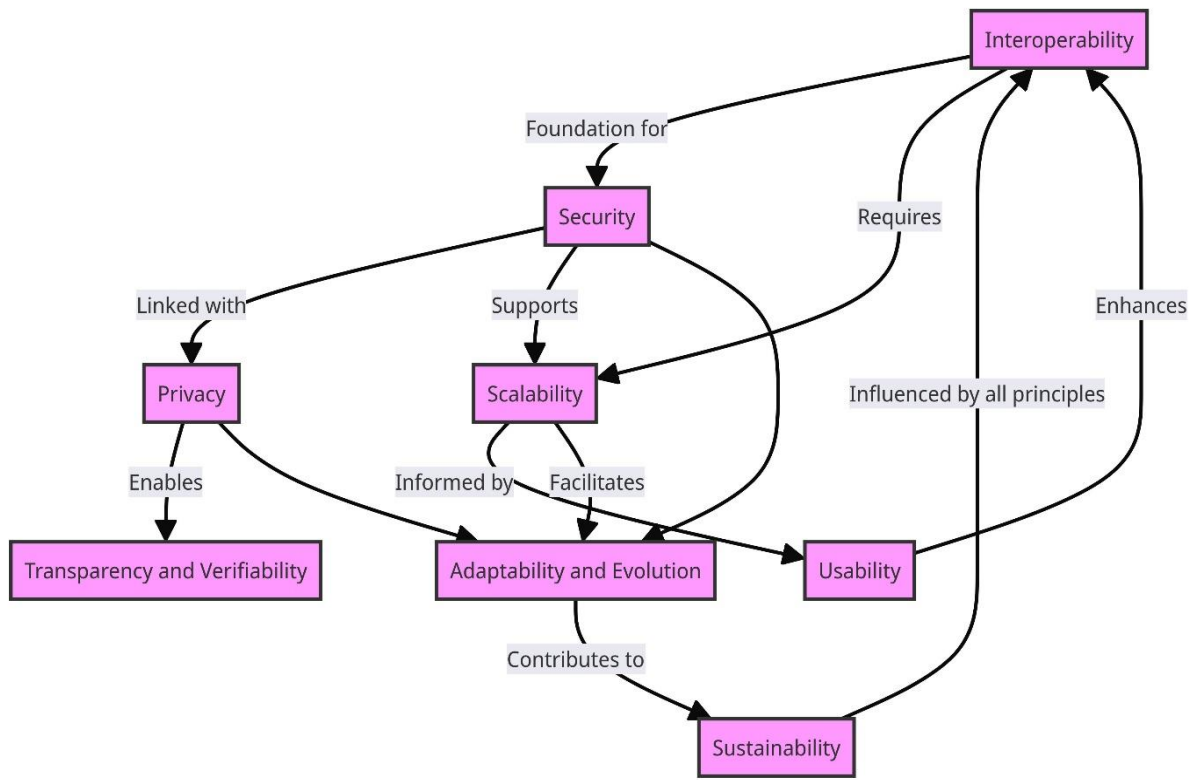

**Figure 2: Core Principles of BACIP.**

Figure 2 visualizes the core principles of the Blockchain Academic Credential Interoperability Protocol (BACIP) and how they are interconnected. Each principle is a node, interconnected to reflect the protocol's comprehensive and synergistic structure. 'Interoperability' sits at the core, linked closely with 'Security' and 'Privacy', which then lead to 'Transparency and Verifiability'. 'Scalability' is essential for growth, supported by 'Usability' for ease of access, while 'Adaptability and Evolution' ensure future relevance. 'Sustainability' represents the commitment to long-term viability. Together, these principles ensure a robust and adaptable framework for academic credential verification.

### 4.4 OPERATIONAL COMPONENTS

The phase of evaluating and selecting operational components for BACIP involved a detailed analysis of cutting-edge studies in the field of academic credential verification using blockchain technology. Below are the studies considered and the key elements identified for integration into BACIP:

#### 4.4.1 BCERT

In the study "BCERT - A Decentralized Academic Certificate System Distribution Using Blockchain Technology" by Elva Leka and Besnik Selimi (2020), the authors propose a novel system leveraging blockchain technology for the storage, distribution, and verification of academic certificates. Employing Ethereum smart contracts alongside the InterPlanetary File System (IPFS) for decentralized file storage, BCert aims to enhance both the efficiency and security of academic credential management. This approach addresses significant issues associated with traditional academic certificates, such as forgery and verification inefficiencies, by harnessing blockchain's core features: decentralization, immutability, and transparency.

When considering the integration of components from this study into the Blockchain Academic Credential Interoperability Protocol (BACIP), several elements could strengthen the protocol:

Smart Contracts for Credential Issuance and Management: Implementing smart contracts would automate and secure the processes of issuing, managing, and verifying academic credentials, ensuring their authenticity and integrity.

IPFS for Credential Storage: Incorporating IPFS would provide a decentralized, efficient, and scalable storage solution for academic credentials, ensuring their long-term availability and durability.

Data Encryption with AES: Employing the AES algorithm for data encryption before its transaction on the blockchain would ensure the confidentiality and protection of students' personal and academic information.

These components align BACIP with principles of interoperability, security, privacy, and transparency and open up new possibilities for quick and reliable credential verification on a global scale. Adopting this technology could significantly transform the higher education landscape, offering a more reliable and accessible system for academic credential management.

### 4.4.2 Blockchain for Global Education

In the comprehensive study "Blockchain for Global Education" by Lakshmana Kumar Ramasamy and Firoz Khan (2024), the authors explore the multifaceted applications of blockchain technology in revolutionizing the educational sector. Addressing challenges such as credential verification, resource distribution, and cross-border academic mobility, the study highlights blockchain's potential in creating a more secure, transparent, and accessible educational environment. Through the examination of decentralized databases, digital identities, e-learning platforms, and global credit transfers, Ramasamy and Khan showcase how blockchain can solve enduring issues in education.

When considering the integration of concepts from this study into the Blockchain Academic Credential Interoperability Protocol (BACIP), the following components could be instrumental:

Global Credit Transfer System: This component, inspired by the study's discussion on facilitating seamless cross-border academic mobility, would leverage blockchain to create a universally recognized system for credit transfer. By ensuring that academic credits are verifiable and transferable across borders, this system would enhance BACIP's principle of interoperability and accessibility in global education.

Decentralized Educational Resource Marketplace: Drawing from the study's exploration of a blockchain-based marketplace for educational resources, this component would enable the sharing, accessing, and monetizing of educational content in a secure and transparent manner. By utilizing smart contracts for rights management and transactions, this marketplace aligns with BACIP's goals of transparency and accessibility, ensuring equitable distribution of educational materials.

Student Digital Identity Management: Reflecting the study's insights on the importance of secure digital identities for students, this component would implement blockchain to provide students with a secure and portable digital identity. This identity would facilitate access to and management of their academic records, supporting BACIP's principles of privacy and security while promoting student autonomy over personal data.

Decentralized E-Learning Platform: Inspired by the study's vision for blockchain-enhanced e-learning systems, this component would offer a decentralized platform for online education. It aims to address current limitations in e-learning by ensuring data privacy, content integrity, and personalized learning paths through blockchain technology, thereby enhancing BACIP's usability and adaptability in digital education.

### 4.4.3 Incorporating Cryptographic-based Blockchain Technology for Degree Automation

In the study "Incorporating a Cryptographic-based Blockchain Technology to Revolutionize Degree Automation" by Ouadoud Oumaima, Tarik Chafiq, Mohammed Ouadoud, and Ahmed Eddaoui (2023), the focus is on the digitization and automation of university diplomas and certificates using blockchain technology. This approach aims to overcome the limitations of traditional methods, which rely on trusted third parties like universities, by providing a sustainable, secure, and universally valued solution for degree management. The authors propose a model that combines public and private blockchain technologies, specifically Ethereum for public transactions and Hyperledger for private, confidential transactions, to manage the entire lifecycle of academic credentials.

This study addresses the core challenges in academic credential management: security, verification efficiency, and international accreditation. By leveraging blockchain's decentralization, immutability, and transparency, the proposed model aims to streamline the verification process, enhance security against forgery, and improve the portability and recognition of credentials across borders.

When considering the integration of concepts from this study into the Blockchain Academic Credential Interoperability Protocol (BACIP), the following components emerge as potential enhancements to the protocol:

Dual Blockchain Architecture: Implementing a hybrid model that utilizes both public (Ethereum) and private (Hyperledger) blockchains. This structure would allow BACIP to offer both transparency and privacy where needed, catering to the diverse needs of academic credential management.

Automated Credential Management: The study's approach to automating the issuance, verification, and management of diplomas through smart contracts could significantly streamline BACIP's operations, ensuring efficiency and reducing reliance on manual, error-prone processes.

Enhanced Security and Verification: Incorporating cryptographic techniques for the secure signing and verification of academic credentials. This would bolster the integrity and trustworthiness of credentials managed through BACIP, addressing concerns around forgery and unauthorized alterations.

International Accreditation and Portability: By adopting a model designed for global recognition, BACIP could facilitate smoother cross-border education pathways and career mobility for students, ensuring that credentials are easily verifiable and valued internationally.

Considerations for Integration:

Scalability and Cost Management: The hybrid blockchain model must be scalable to accommodate a global educational ecosystem while managing transaction costs effectively, especially on the public blockchain.

Privacy and Data Protection: Ensuring compliance with international data protection regulations (e.g., GDPR) when handling sensitive student information on a decentralized network.

Stakeholder Engagement: Collaborating with educational institutions, accreditation bodies, and technology providers to ensure the system meets the diverse needs and standards of global education.

This analysis concludes that "Incorporating a Cryptographic-based Blockchain Technology to Revolutionize Degree Automation" provides valuable insights and frameworks that could significantly contribute to the development of BACIP, aligning with its objectives to modernize, secure, and internationalize academic credential management. Incorporating these components would position BACIP at the forefront of educational innovation, leveraging blockchain technology to address current challenges in credential verification and management on a global scale.

### 4.4.4 Building Smart Contract-Based Higher Education Systems Using Zero-Knowledge Ethereum Virtual Machine

In their study, "Toward Building Smart Contract-Based Higher Education Systems Using Zero-Knowledge Ethereum Virtual Machine," Dénes László Fekete and Attila Kiss (2023) investigate the application of blockchain technology to enhance the issuance, storage, and verification of higher education certificates. The study advocates for a permissionless, verifiable education system leveraging the latest advancements in Ethereum Virtual Machine (EVM) technology, specifically zero-knowledge proofs (zkEVM), to ensure a decentralized, secure, and privacy-preserving method of managing academic credentials.

This approach addresses critical issues in the current academic credential management system, including inefficiency, vulnerability to fraud, and lack of global interoperability. By utilizing a combination of public and private blockchain technologies, the proposed system aims to provide a sustainable solution that enhances the security, transparency, and accessibility of academic credentials.

Integrating concepts from this study into the Blockchain Academic Credential Interoperability Protocol (BACIP) could lead to the following potential enhancements:

Zero-Knowledge Proof for Privacy and Security: Implementing zkEVM within BACIP could ensure that academic credentials are verified without revealing any underlying personal information, thereby enhancing privacy and security. This aligns with BACIP's principles of security and privacy.

Hybrid Blockchain Architecture: The use of both public and private blockchains could allow BACIP to optimize the balance between transparency and confidentiality, ensuring that academic credentials are universally verifiable while protecting sensitive student information.

Permissionless Verifiable Education System: Adopting a permissionless approach would lower the trust assumptions required between entities, making BACIP more robust against attacks and reducing the potential for malicious behavior.

Modular Blockchain Structure for Scalability: The recommendation of a modular blockchain structure implies that BACIP could incorporate various functionalities and capabilities from the examined works into one cohesive system. This would address the principles of scalability and adaptability by enabling BACIP to evolve over time and integrate additional features as necessary.

Considerations for Integration:

Ensuring GDPR Compliance: Given the emphasis on privacy and data protection within the EU, any implementation of BACIP must be compliant with the General Data Protection Regulation (GDPR), especially when handling personal data on a blockchain.

Balancing Decentralization and Efficiency: While pursuing a decentralized approach for credential verification, it's critical to maintain system efficiency and user-friendliness to encourage adoption by educational institutions and students.

Stakeholder Collaboration: For the successful deployment of BACIP, collaboration with a wide range of stakeholders, including educational institutions, technology providers, and regulatory bodies, will be essential to ensure the system meets diverse needs and legal requirements.

Integrating these components from Fekete and Kiss's study into BACIP could significantly contribute to its objective of revolutionizing the management of academic credentials through blockchain technology. By embracing zero-knowledge proofs and a hybrid blockchain architecture, BACIP can advance towards a more secure, private, and interoperable system for credential management across the globe.

### 4.4.5 Verification of Education Credentials on European Blockchain Services Infrastructure (EBSI)

In "Verification of Education Credentials on European Blockchain Services Infrastructure (EBSI): Action Research in a Cross-Border Use Case between Belgium and Italy," by Evrim Tan et al. (2023), the authors present a pioneering application of blockchain technology for verifying education credentials across borders. The study explores the implementation of the EBSI, a major initiative by the EU and European Blockchain Partnership, aimed at leveraging blockchain technology to create cross-border services for public administrations, businesses, citizens, and their ecosystems. This research emphasizes the use of digital wallets, verifiable credentials, and decentralized identifiers to facilitate secure and interoperable academic mobility and service verification.

Key Components for BACIP Integration:

EBSI-Compliant Digital Wallets and Verifiable Credentials: The study highlights the implementation of digital wallets and verifiable credentials compliant with the EBSI framework. For BACIP, integrating similar technologies could enhance the secure issuance, storage, and verification of academic credentials, ensuring their acceptance and recognition across the EU.

Decentralized Identifiers for Academic Credentials: Utilizing decentralized identifiers as part of the blockchain protocol can significantly improve the management and verification of academic credentials. This approach aligns with BACIP's goals to facilitate secure, tamper-proof, and easily verifiable academic records.

Cross-Border Verification of Education Credentials: The research provides insights into the technical and institutional requirements for the cross-border verification of education credentials. For BACIP, incorporating these findings could address interoperability challenges and support the seamless exchange of academic information between institutions in different countries.

Challenges and Solutions for Wider Adoption: The pilot identifies key challenges for the wider adoption of blockchain in credential verification, including onboarding governance, data scheme issuance for transcript validation, and interoperability issues. BACIP can leverage these insights to anticipate and mitigate similar challenges, focusing on developing a user-friendly, accessible, and widely adopted protocol.

Considerations for Integration:

Scalability and Efficiency: Ensuring that BACIP can handle a high volume of credential verifications efficiently is crucial for its success. Lessons from the EBSI implementation regarding system performance and scalability should inform the technical design of BACIP.

Legal and Regulatory Compliance: Adhering to legal frameworks and data protection laws (e.g., GDPR) is essential. BACIP should incorporate privacy-preserving features and comply with regulatory requirements to facilitate its adoption.

Stakeholder Collaboration: Collaborating with educational institutions, government bodies, and technology providers across the EU will be key to overcoming implementation challenges and ensuring that BACIP meets the diverse needs of its users.

The findings from this EBSI pilot study offer valuable insights into the practical application of blockchain for credential verification in an international context. By addressing the identified challenges and integrating key components from the study, BACIP can significantly advance its development towards creating a secure, interoperable, and efficient protocol for academic credential management.

## 4.5 SELECTED OPERATIONAL COMPONENTS

In synthesizing the final selection of operational components for the Blockchain Academic Credential Interoperability Protocol (BACIP), the culmination of research, analysis, and refinement has resulted in the integration of pivotal elements. These components, drawn from leading-edge studies in blockchain technology's application to academic credential verification, have been meticulously aligned with the core principles established for BACIP. This integration enhances BACIP's functionality and ensures adherence to its foundational pillars: interoperability, security, privacy, and more.

Each component, from the adoption of smart contracts for credential management to the employment of decentralized storage solutions and beyond, has been deliberately selected to address the dynamic challenges and requirements of modern academic credential verification. The incorporation of these operational elements reflects a careful consideration of scalability, ensuring that as BACIP grows, it remains efficient and secure, capable of handling an increasing volume of transactions and users without sacrificing performance.

The introduction of a global credit transfer system decentralized educational resource marketplace, and the management of student digital identities through blockchain technology signifies a leap forward in how educational achievements and resources are exchanged and authenticated worldwide. The use of dual blockchain architecture and cryptographic techniques for the secure signing and verification of academic credentials underscores BACIP's commitment to maintaining integrity and fostering trust within the academic sphere.

The assembly of these components into BACIP not only symbolizes a significant stride in academic credential verification but also serves as a template for future technological advancements within the educational sector. These operational components of BACIP, foundational to its deployment and efficacy, are illustrated in **Figure 3**, offering a visual representation of how each contributes to the protocol's comprehensive structure and overarching vision.

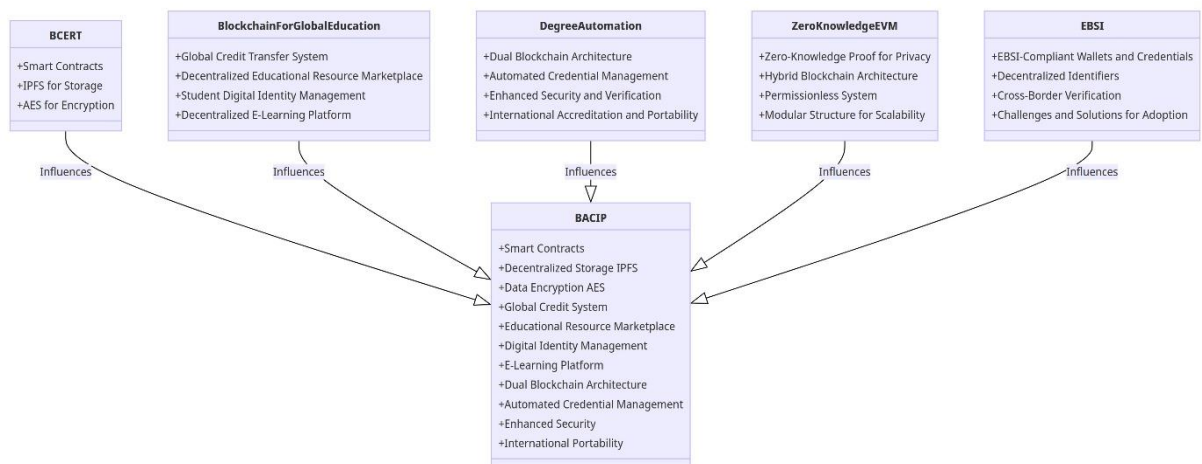

**Figure 3: Operational Components of BACIP and Their Influences.**

Figure 3 presents the chosen operational components that have been integrated into the architecture of BACIP, emphasizing how each contributes to the protocol's functionality. Features like smart contracts, decentralized storage, and global credit transfer systems are depicted as foundational elements that bolster a resilient structure. This visualization provides a concise view of how each component, derived from leading research, fortifies BACIP, underscoring the commitment to integrating advancements that enhance security, privacy, and scalability in academic credential verification.

### 4.5.1 Smart Contracts for Credential Issuance and Management
Utilize blockchain smart contracts to automate the processes of issuing, managing, and verifying academic credentials. These contracts ensure the authenticity, integrity, and non-repudiation of academic records.

### 4.5.2 Decentralized Storage Solutions (IPFS)
Implement the InterPlanetary File System (IPFS) or similar decentralized storage technologies to store credential data. This enhances data availability and durability while ensuring efficiency in data retrieval.

### 4.5.3 Data Encryption with AES
Employ Advanced Encryption Standard (AES) for encrypting sensitive data before recording it on the blockchain. This practice safeguards students' personal and academic information against unauthorized access.

### 4.5.4 Global Credit Transfer System
Develop a system within BACIP that leverages blockchain to facilitate seamless cross-border academic mobility. This system will enable the verifiable and transferable recognition of academic credits across different educational institutions globally.

### 4.5.5 Decentralized Educational Resource Marketplace
Create a blockchain-based marketplace for educational resources that allows for secure sharing, accessing, and monetizing educational content. This marketplace utilizes smart contracts for rights management and transactions, promoting equitable distribution of educational materials.

### 4.5.6 Student Digital Identity Management

Incorporate blockchain technology to provide students with secure and portable digital identities. These identities facilitate efficient access to and management of academic records, enhancing students' control over their personal data.

### 4.5.7 Decentralized E-Learning Platform

Offer a decentralized platform for online education that addresses current limitations by ensuring data privacy, content integrity, and personalized learning paths through blockchain technology.

### 4.5.8 Dual Blockchain Architecture

Adopt a hybrid model that utilizes both public and private blockchain technologies to cater to the diverse needs of academic credential management. This model combines the transparency of public blockchains with the privacy features of private blockchains.

### 4.5.9 Automated Credential Management Through Smart Contracts

Streamline the issuance, verification, and management of diplomas and other academic documents through smart contracts, reducing reliance on manual, error-prone processes.

### 4.5.10 Enhanced Security and Verification Using Cryptographic Techniques

Integrate cryptographic techniques to securely sign and verify academic credentials, bolstering the integrity and trustworthiness of credentials managed through BACIP.

### 4.5.11 International Accreditation and Portability

Design BACIP to support the global recognition of academic credentials, facilitating smoother education pathways and career mobility for students worldwide.

## 4.6 STANDARDIZATION AND TECHNICAL SPECIFICATIONS

This section focuses on Phase III of the development of the Blockchain Academic Credential Interoperability Protocol (BACIP), specifically on the standardization and technical definition of its components. During this phase, the technical specifications essential for the operability and functionality of the protocol were precisely established and documented.

The process included the evaluation and selection of network architectures, data models, consensus protocols, and security mechanisms, all aligned with the established principles of interoperability, security, and privacy. The applied methodology ensured that each technical component not only integrated efficiently within the BACIP ecosystem but also addressed the specific challenges of academic credential verification.

A rigorous review of relevant literature and empirical studies was conducted to support the technical decisions and ensure that the adopted solutions were both innovative and evidence based. This methodological approach allowed for detailed documentation of the technical specifications underpinning the BACIP protocol, highlighting the importance of a solid technical foundation for the success of any credential verification system.

### 4.6.1 Digital Certificates Design

**Unique Identifier (ID)**

A universally unique identifier (UUID) is crucial in distinguishing each credential uniquely and unequivocally. The version 4 UUID is randomly generated, making it an optimal choice for ensuring that each ID is unique without the need for a central registry, which aligns well with the decentralized nature of blockchain technology.

**Technical Implementation:**

UUID v4 is generated based on random numbers, but for an additional layer of security and traceability, combining this with a timestamp and issuer-specific information further fortifies the ID's uniqueness and makes it traceable. The cryptographic algorithm used for this process ensures that the IDs, once generated, cannot be predicted or replicated, upholding the integrity and confidentiality of the credentialing system.

**Literature Support:**

Research highlights the importance of unique identifiers in the management of digital data and specifically in blockchain applications where data integrity and non-repudiation are critical (Nakamoto, 2008; Swan, 2015). By leveraging a cryptographic method, the UUID can integrate additional security layers that make the credentialing system robust against various security threats, such as impersonation or duplication (Buterin, 2014).

**Relevance to Blockchain:**

In the context of blockchain, the use of such a UUID aligns with the technology's principles of decentralization and transparency. Each credential can be verified independently without the need for an intermediary, thus simplifying the verification process and enhancing the trust in the digital certificates issued.

**Example:**

A UUID v4 might be generated as follows: 123e4567-e89b-12d3-a456-426614174000

When combined with timestamp and issuer information, it ensures that each ID is not only unique but also carries metadata that can further assist in auditing and compliance processes, essential for academic credentials.

**Standard Metadata**

The adoption of JSON-LD (JavaScript Object Notation for Linked Data) for structuring metadata is crucial for ensuring semantic interoperability between different systems and platforms. JSON-LD facilitates the integration of linked data in a way that is both machine-readable and human-understandable, easing data integration and reuse across various applications.

**Technical Implementation:**

JSON-LD is a lightweight specification for structuring related and linked data. It allows the definition of contexts that map JSON object properties to concepts in web ontologies, making the data globally understandable. Within the context of digital credentials, metadata fields can include:

issuer: URI identifying the issuer.

recipient: Object detailing the recipient, including a DID (Decentralized Identifier).

credentialSubject: Description of the credential, following the Schema.org schema for educational credentials.

issueDate and expirationDate: Dates formatted according to ISO 8601.

signature: Object containing the type of digital signature and the signature itself, using algorithms like ES256 for ECDSA on the P-256 curve.

**Literature Support:**

The adoption of JSON-LD is supported by its capability to enhance data interoperability across different platforms, a crucial advantage in educational and academic settings where credentials need to be recognized and verified by multiple entities (W3C, 2020). Additionally, studies such as those by Christidis and Devetsikiotis (2016) highlight that metadata standardization is essential to ensure the validity and authenticity of digital credentials on the blockchain.

**Relevance to Blockchain:**

In blockchain applications, using JSON-LD facilitates the integration of credentials across different blockchains and systems without losing data coherence. This is particularly valuable in higher education applications where credentials like diplomas and certificates need to be interchangeable and verifiable globally.

**Example:**

An example of how JSON-LD might be structured for a digital credential is:

```
{
  "@context": "https://schema.org",
  "@type": "EducationalOccupationalCredential",
  "issuer": "https://university.example.edu",
  "recipient": {
    "type": "Person",
    "id": "did:example:abcdef",
```

```
    "name": "Juan Pérez"
  },
  "credentialSubject": {
    "degree": "MSc Computer Science"
  },
  "issueDate": "2021-05-01",
  "expirationDate": "2026-05-01",
  "signature": {
    "type": "ES256",
    "signatureValue": "base64EncodedSignatureValueHere"
  }
}
```

**Signature Algorithms**

The integrity and authenticity of digital certificates are paramount, particularly in blockchain applications where the validity of academic credentials must be unquestionable. Signature algorithms play a crucial role in ensuring that digital certificates have not been altered post-issuance and that they are verifiably issued by a legitimate entity.

**Technical Implementation:**

The BACIP recommends using ECDSA with the P-256 curve, commonly known as ES256, due to its balance of security and computational efficiency. This choice is widely accepted in various security protocols and ensures broad compatibility with existing and emerging technologies.

Additionally, to accommodate advancements in cryptographic practices and address potential future security challenges, support for EdDSA, specifically Ed25519, is proposed. Ed25519 is renowned for its fast signature verification, low computational overhead, and increased resistance to side-channel attacks, making it an attractive option for systems where performance and security are critical.

**Literature Support:**

The selection of ECDSA and EdDSA is supported by their widespread adoption in security-sensitive applications, as documented in studies and recommendations by cryptographic authorities (Bernstein and Lange, 2017; Johnson et al., 2001). Their applicability in blockchain technologies is particularly noted for their robustness in securing transactions and maintaining the integrity of distributed ledgers.

**Relevance to Blockchain:**

In blockchain systems, the choice of signature algorithm directly impacts the trust model and the efficiency of the network. These algorithms ensure that all transactions recorded on the blockchain—such as the issuance, transfer, or revocation of digital certificates—are immutable and traceable to verified issuers, thereby upholding the system's credibility and reliability.

**Example:**

In practical terms, a digital certificate might include a signature object like this:

```
"signature": {
  "type": "ES256",
  "signatureValue": "Base64EncodedSignature"
}
```

Here, ES256 indicates the use of ECDSA with the P-256 curve. For systems requiring higher security assurances or faster processing times, Ed25519 might be utilized, enhancing the protocol's flexibility and security stance.

This focus on signature algorithms will help ensure that your digital certificate system in BACIP remains secure, scalable, and adaptable to future technological developments and potential cryptographic vulnerabilities.

**Revocation Mechanisms**

The ability to revoke digital certificates is critical in maintaining the security and relevance of credentials. Revocation might be necessary due to various reasons, such as the compromise of the issuing authority's security, the certificate holder's change in status, or expiration of the certificate's intended purpose. Effective revocation mechanisms ensure that all stakeholders can trust the current validity of any given credential.

**Technical Implementation:**

The implementation of revocation mechanisms in BACIP could involve a dedicated smart contract on the blockchain that manages the revocation list. Each digital certificate can be associated with a unique identifier, which can be added to this revocation list if the certificate is deemed invalid. Utilizing blockchain for this purpose ensures that once a certificate is revoked, this status is immutable and transparent to all participants in the network.

Alternatively, a more dynamic approach could be the use of a distributed ledger to record and check the real-time status of each certificate without actually storing the entire certificate on the blockchain. This method enhances privacy and efficiency by only querying for the status based on need, using cryptographic proofs to ensure data integrity.

**Literature Support:**

Studies on digital certificate management emphasize the importance of robust revocation mechanisms to prevent the misuse of invalid credentials and maintain system integrity (Kohnfelder, 1978). Blockchain technology offers a decentralized and transparent way to manage revocations, ensuring all network participants can trust the system's updates without relying on a single authority (Narayanan et al., 2016).

**Relevance to Blockchain:**

Blockchain's inherent characteristics of immutability and transparency make it an ideal platform for implementing revocation mechanisms. By leveraging smart contracts or ledger-based solutions, the system can instantly propagate revocation status across the network, ensuring that all parties have access to up-to-date and accurate certificate validity information.

**Example:**

Here is a simplified example of how a revocation mechanism might be represented in a smart contract:

```solidity
pragma solidity ^0.5.0;

contract RevocationRegistry {
    mapping(bytes32 => bool) private revokedCertificates;

    // Event that is emitted when a certificate is revoked
    event CertificateRevoked(bytes32 indexed certId);

    // Function to revoke a certificate
    function revokeCertificate(bytes32 certId) public {
        revokedCertificates[certId] = true;
        emit CertificateRevoked(certId);
    }

    // Function to check if a certificate is revoked
    function isRevoked(bytes32 certId) public view returns (bool) {
        return revokedCertificates[certId];
    }
```

```
}
```

This feature ensures that any entity within the network can verify the revocation status of a certificate in real-time, thereby upholding the trust and security of the digital credential system.

### 4.6.2 Communication Protocols

**RESTful API**

RESTful APIs are crucial for ensuring that distributed systems like blockchain networks can communicate effectively and efficiently. They facilitate standardized access to web services, which is essential for the interoperability of different systems and platforms.

**Technical Implementation:**

The BACIP defines specific RESTful endpoints such as /issueCredential, /verifyCredential, and /revokeCredential. Each endpoint is designed to handle JSON formatted data, which simplifies data interchange and ensures consistency across the network.

**Literature Support:**

Fielding and Taylor (2002) describe the REST architectural style in their seminal work, emphasizing its scalability, generality, and independence from resources and mechanisms, making it suitable for blockchain applications where such traits are highly valued.

**Relevance to Blockchain:**

In blockchain systems, RESTful APIs facilitate the integration of blockchain data with other systems and web interfaces, allowing for more complex applications and services. They support the blockchain's decentralized nature by allowing various clients to interact with the network without needing direct access to the blockchain ledger.

Example:

```
POST /issueCredential HTTP/1.1
Host: api.university.example.edu
Content-Type: application/json
Content-Length: ...

{
  "issuer": "https://university.example.edu",
```

```
  "recipient": {
    "name": "John Doe",
    "id": "did:example:123"
  },
  "credential": {
    "type": "Diploma",
    "course": "BSc Computer Science"
  }
}
```

**Data Exchange Format**

Uniform data exchange formats are essential for ensuring that information is consistently understood across all system interactions. JSON is particularly advantageous due to its compatibility with web technologies and its ability to be parsed easily by various programming languages.

**Technical Implementation:**

JSON schemas define the structure of data for API requests and responses. This ensures that all data interactions follow a standardized format, reducing errors and improving system reliability.

**Literature Support:**

Bray (2017) highlights the advantages of using JSON for data interchange, citing its easy integration into existing systems and its widespread support across programming environments.

**Relevance to Blockchain:**

The use of JSON in blockchain applications ensures that data exchanged between nodes and external systems is in a format that is readily accessible and easily integrable. This facilitates the seamless transfer and verification of data across the network, crucial for maintaining transparency and integrity in blockchain operations.

**Example:**

```
{
  "$schema": "http://json-schema.org/draft-07/schema#",
  "type": "object",
  "properties": {
    "issuer": {"type": "string", "format": "uri"},
```

```
      "recipient": {
        "type": "object",
        "properties": {
          "name": {"type": "string"},
          "id": {"type": "string"}
        },
        "required": ["id"]
      },
      "credential": {
        "type": "object",
        "properties": {
          "type": {"type": "string"},
          "course": {"type": "string"}
        },
        "required": ["type", "course"]
      }
   },
   "required": ["issuer", "recipient", "credential"]
}
```

### 4.6.3 Consensus and Validation Mechanisms

**Consensus Mechanism**

The consensus mechanism is a fundamental aspect of blockchain technology, ensuring that all participating nodes in the network agree on the single, truthful state of the ledger. This prevents any single entity from controlling or altering the recorded data unilaterally, which is crucial for a decentralized system like BACIP.

**Technical Implementation:**

BACIP employs the Istanbul Byzantine Fault Tolerant (IBFT) consensus mechanism. IBFT improves upon classic Byzantine Fault Tolerance (BFT) by ensuring transaction finality without forks within permissioned blockchain networks. This mechanism provides fast block times and efficient decision-making processes, crucial for academic environments where timely and accurate updates to credentials are necessary.

**Literature Support:**

The selection of IBFT is supported by its proven stability and efficiency in network operations as discussed by Sousa et al. (2018), who highlight its suitability for networks requiring robustness against adversarial conditions within a permissioned consortium.

**Relevance to Blockchain:**

In blockchain systems, employing IBFT aligns with the need for high security, speed, and fault tolerance. It supports the integrity and transparency of the blockchain, ensuring that all credentials issued, verified, or revoked are accepted and recognized by all network participants without discrepancies.

**Example:**

In an IBFT network, each node participates in the consensus process by proposing blocks and validating transactions through a multi-stage agreement protocol. This involves preparing, committing, and finalizing blocks, ensuring no conflicting transactions (double-spends, invalid data) are recorded.

**Transaction Validation Process**

Validating transactions in a blockchain is essential to prevent fraud and ensure the network operates correctly. Each transaction must be rigorously checked to ensure it conforms to the network's rules and does not attempt to corrupt the database.

**Technical Implementation:**

In BACIP, transaction validation is executed by validator nodes, which check each transaction against the network's protocol rules. This includes verifying digital signatures, ensuring the credentials have not been revoked, and confirming the transaction's compliance with the network's established criteria.

**Literature Support:**

As per Castro and Liskov (2002), implementing a rigorous transaction validation process is vital for maintaining the security and functionality of the blockchain, preventing malicious actors from exploiting system vulnerabilities.

**Relevance to Blockchain:**

The validation process is integral to maintaining the blockchain's integrity. By ensuring that each transaction is legitimate and conforms to established rules, BACIP safeguards the network against fraudulent activities and ensures that the credential data remains accurate and trustworthy.

Authentication: Use of JWT (JSON Web Tokens) to authenticate API requests, signed with issuers' private keys.

Authorization and Access Control: Implementation of smart contracts to manage roles (issuer, verifier, student) and associated permissions, using ACLs to control access to operations on the blockchain.

**Example:**

A validator node receives a transaction issuing a new credential. It checks:

The digital signature against the issuer's public key.

The credential's status in the revocation list.

The compliance of the credential's format with the predefined schema.

### 4.6.4 Identity and Access Management

**Authentication**

In the blockchain environment, particularly in an academic setting where credentials are issued and verified, robust authentication mechanisms are essential to ensure that only authorized entities can perform specific actions. Authentication provides a security layer that confirms the identity of users or nodes before they can make transactions or access sensitive information.

**Technical Implementation:**

BACIP utilizes JSON Web Tokens (JWT) to manage authentication. JWTs are signed using a secure method (as previously discussed, such as ES256 or Ed25519) and contain claims that specify the identity of the user and the permissions that the user has. This token-based system allows for stateless authentication, which is ideal for distributed systems like blockchain.

**Literature Support:**

According to Jones and Hildebrand (2015), JWTs offer an efficient and secure way to assert claims between two parties, making them highly suitable for applications requiring clear, verifiable identities.

**Relevance to Blockchain:**

Authentication via JWTs aligns well with the decentralized and distributed nature of blockchain. It ensures that actions performed on the network can be traced back to verified identities, enhancing the network's security and integrity.

**Example:**

A typical JWT might include:

```
{
  "alg": "ES256",
```

```
  "typ": "JWT"
}
{
  "sub": "issuer123",
  "name": "University of Blockchain",
  "iat": 1516239022,
  "role": "Issuer"
}
```

This token is then used in API requests to authenticate actions such as issuing or revoking credentials.

**Authorization and Access Control**

Authorization determines what an authenticated user is permitted to do within the system. Effective access control mechanisms are necessary to enforce policy decisions that restrict users' actions based on their roles and the permissions associated with these roles.

**Technical Implementation:**

BACIP implements smart contracts to manage roles and permissions. Access control lists (ACLs) are used to define and enforce what actions each role (issuer, verifier, student) can perform on the blockchain. This setup ensures that operations such as issuing, verifying, or revoking credentials are performed only by authorized entities.

**Literature Support:**

As suggested by Pretschner et al. (2006), access control mechanisms in decentralized systems like blockchain must be robust and adaptable to ensure that they cater to evolving network and security requirements.

**Relevance to Blockchain:**

The use of smart contracts for access control leverages the inherent security and transparency features of the blockchain, ensuring that authorization policies are not only enforced but also immutable and auditable by all network participants.

**Example:**

A smart contract might manage permissions as follows:

```
// Solidity pseudocode
contract AccessControl {
```

```
    mapping(address => uint) public rolePermissions;

    function setPermissions(address user, uint role) public {
        require(msg.sender == admin);
        rolePermissions[user] = role;
    }

    function issueCredential(address user) public {
        require(rolePermissions[user] & ISSUE_PERMISSION);
        // issue credential
    }
}
```

This contract ensures that only users with the correct permissions can issue credentials.

### 4.6.5 Privacy and Security Standards
**Data Encryption**

Protecting sensitive data stored on the blockchain is paramount, especially when dealing with personal and academic information that requires confidentiality. Data encryption ensures that even if data is accessed, it remains unreadable and secure without the appropriate decryption keys.

**Technical Implementation:**

BACIP uses AES-GCM with 256-bit keys for encrypting sensitive data before it is stored on the blockchain. This symmetric encryption standard offers both security and performance, providing strong encryption quickly, which is essential for maintaining the blockchain's performance.

**Literature Support:**

According to Dworkin (2007), AES-GCM is recommended by the National Institute of Standards and Technology (NIST) for its robustness and efficiency, making it an excellent choice for secure data transmission and storage in systems requiring high throughput and security.

**Relevance to Blockchain:**

In blockchain, encryption enhances data privacy and security significantly. While the blockchain is inherently secure due to its decentralized nature and cryptographic underpinnings, encrypting data adds an additional layer of security, protecting against potential threats or data breaches.

**Example:**

This Python code snippet demonstrates encryption and decryption of data using the AES-GCM encryption algorithm:

```python
from Crypto.Cipher import AES
from Crypto.Random import get_random_bytes
# Function to encrypt data
def encrypt_data(data, key):
    cipher = AES.new(key, AES.MODE_GCM)
    ciphertext, tag = cipher.encrypt_and_digest(data.encode('utf-8'))
    return ciphertext, cipher.nonce, tag
# Function to decrypt data
def decrypt_data(ciphertext, key, nonce, tag):
    cipher = AES.new(key, AES.MODE_GCM, nonce=nonce)
    plaintext = cipher.decrypt_and_verify(ciphertext, tag)
    return plaintext.decode('utf-8')
# Encryption key (must be secure and kept secret)
encryption_key = get_random_bytes(32)  # 256-bit key
# Data to encrypt
data_to_encrypt = "Password: mySecurePassword123"
# Encrypt the data
encrypted_data, nonce, tag = encrypt_data(data_to_encrypt, encryption_key)
# Simulate storing in the blockchain (in real life, this would be part of a transaction)
blockchain_storage = {
    'encrypted_data': encrypted_data,
    'nonce': nonce,
    'tag': tag
}
# Simulate data retrieval from the blockchain
retrieved_encrypted_data = blockchain_storage['encrypted_data']
```

```
retrieved_nonce = blockchain_storage['nonce']
retrieved_tag = blockchain_storage['tag']
# Decrypt the retrieved data
decrypted_data    =    decrypt_data(retrieved_encrypted_data,
encryption_key, retrieved_nonce, retrieved_tag)
print("Decrypted data:", decrypted_data)
```

The code imports necessary libraries from the Crypto package, specifically AES for encryption and get_random_bytes for generating a random encryption key. It defines an encrypt_data function to encrypt data using AES-GCM encryption, producing ciphertext, a nonce, and a tag. Similarly, a decrypt_data function decrypts data using AES-GCM decryption. A 256-bit encryption key is generated using get_random_bytes. The example data "Password: mySecurePassword123" is encrypted, and the resulting encrypted data, nonce, and tag are stored in simulated blockchain storage. Upon retrieval, the encrypted data, nonce, and tag are decrypted using the decrypt_data function. Finally, the decrypted data is printed to the console, illustrating a basic implementation of data encryption and decryption using Python with AES-GCM encryption for securing data in a simulated blockchain environment.

**Data Rights**

Given the global nature of academic credentials and the sensitive information they contain, it is crucial for BACIP to comply with a variety of educational regulations and privacy laws, such as FERPA in the United States, GDPR in Europe, and other national or regional regulations. Compliance ensures the legality and acceptability of the blockchain system's operations across borders.

**Technical Implementation:**

BACIP incorporates regulatory compliance checks into both the design of its smart contracts and its operational procedures. This includes mechanisms for data protection, such as data minimization, right to erasure ("right to be forgotten"), and explicit consent protocols for data sharing. Moreover, the system can adapt to different regulations by enabling configurable settings for data handling rules that align with specific legal requirements.

**Literature Support:**

Research and case studies on blockchain compliance suggest that integrating legal considerations into blockchain design is not only feasible but essential for widespread adoption, especially in sectors regulated by strict data governance standards (Palmirani et al., 2018).

**Relevance to Blockchain:**

Integrating compliance and regulatory standards directly into the blockchain infrastructure ensures that all transactions and data handling procedures meet the required legal standards. This integration makes BACIP a reliable platform for issuing and managing academic credentials, enhancing trust among institutions, students, and employers worldwide.

**Example:**

```solidity
// Solidity pseudocode for GDPR compliance
contract GDPRCompliance {
    mapping(address => bool) public consentGiven;

    function giveConsent(address user) public {
        consentGiven[user] = true;
    }

    function withdrawConsent(address user) public {
        require(consentGiven[user]);
        consentGiven[user] = false;
        emit ConsentWithdrawn(user);
    }

    function deleteData(address user) public {
        require(consentGiven[user] == false);
        // code to delete user data
        emit DataDeleted(user);
    }
}
```

This contract ensures that the user's consent is managed according to GDPR standards and that data can be deleted upon user request.

### 4.6.6 Interoperability

**Adoption of Open Standards**

Interoperability ensures that diverse systems and technologies can work together seamlessly, which is essential in the context of global education and academic credential verification. Adopting open standards facilitates the integration of blockchain systems with existing educational and technological infrastructures, enhancing the system's utility and reach.

**Technical Implementation:**

BACIP utilizes established open standards such as Decentralized Identifiers (DIDs) and Verifiable Credentials (VCs) from the World Wide Web Consortium (W3C). These standards provide a common framework for creating and managing digital identities and credentials that are platform-independent and can be verified across different systems without the need for centralized control.

**Literature Support:**

The implementation of DIDs and VCs as recommended by the W3C is supported in the literature for enhancing the interoperability and scalability of digital identity systems in decentralized networks (Sporny, Longley, and Chadwick, 2019).

**Relevance to Blockchain:**

Blockchain's decentralized nature makes it an ideal platform for implementing open standards that require robustness, transparency, and security. The use of these standards ensures that credentials stored on the blockchain can be accessed and verified globally, across different sectors and applications, promoting widespread adoption and utility.

**Example:**

```
{
  "@context": "https://www.w3.org/2018/credentials/v1",
  "id": "did:example:123",
  "type": ["VerifiableCredential"],
  "issuer": "did:example:456",
  "issuanceDate": "2020-04-22T11:52:25Z",
  "credentialSubject": {
    "id": "did:example:789",
    "degree": "Bachelor of Science in Blockchain Technology"
  },
  "proof": {
    "type": "RsaSignature2018",
    "created": "2020-04-22T11:52:25Z",
    "proofPurpose": "assertionMethod",
```

```
      "verificationMethod": "https://example.edu/keys/1",
      "jws": "eyJhbGciOiJSUzI1NiIsInR5cCI6IkpXVCJ9..."
  }
}
```

This format allows credentials to be issued and verified across different platforms and systems, making the educational qualifications universally recognized and verifiable.

### 4.6.7 Smart Contract Implementation

**Choice of Solidity**

Solidity is a statically typed programming language designed specifically for developing smart contracts that run on the Ethereum Virtual Machine (EVM). It is the leading language for Ethereum smart contracts and is widely adopted due to its maturity, robust community support, and rich documentation. Choosing Solidity allows leveraging an existing, extensive development ecosystem, which includes tools for testing, deploying, and verifying smart contracts.

**Technical Implementation:**

Solidity's syntax is similar to that of JavaScript, which reduces the learning curve for new developers and enhances the development speed for those familiar with web development. It provides powerful features such as inheritance, libraries, and user-defined types, which are essential for building maintainable and secure smart contracts.

**Literature Support:**

According to Buterin and others (2014), Solidity was designed to cater to the unique needs of blockchain applications, including security features such as static typing and contract-specific primitives that mitigate common contract vulnerabilities.

**Relevance to Blockchain:**

Solidity is specifically crafted for blockchain development, offering constructs that align well with the decentralized and trust-minimized environment of blockchain networks. Its integration into the Ethereum blockchain allows for the implementation of complex business logic directly on the blockchain, ensuring that operations such as the issuance, verification, and revocation of credentials are executed in a decentralized and tamper-proof manner.

**Example:**

Here is a basic example of a Solidity smart contract used for issuing a digital credential:

```
pragma solidity ^0.8.0;
```

```
contract CredentialIssuance {
    struct Credential {
        string degree;
        string issuer;
        address recipient;
    }

    mapping(address => Credential) public credentials;

    function issueCredential(address _recipient, string memory _degree, string memory _issuer) public {
        Credential storage credential = credentials[_recipient];
        credential.degree = _degree;
        credential.issuer = _issuer;
    }
}
```

This contract showcases how to define structures, functions, and storage types in Solidity, focusing on handling credentials efficiently.

### 4.7 OVERVIEW OF BACIP TECHNICAL SPECIFICATIONS

A rigorous standardization and definition of the technical components of the Blockchain Academic Credential Interoperability Protocol (BACIP) is undertaken. This systematic approach is grounded on a critical evaluation of empirical studies and a comprehensive literature review, ensuring that the selection of architectures, data models, smart contracts, consensus mechanisms, and security protocols aligns with the core principles of BACIP. The purpose of this meticulous integration is to incorporate technical components that not only effectively address the unique challenges associated with academic credential verification but also strictly adhere to the guiding principles of the protocol, thereby ensuring its innovative character and structural robustness.

According to **Figure 4**, which illustrates the operational components of BACIP, several key elements are highlighted that contribute directly to the protocol's overall functionality. These elements include smart contracts designed for the issuance and management of credentials, decentralized data storage solutions, and systems designed to facilitate global academic credit transfers. Each component has been selected with the goal of strengthening the overall architecture of BACIP, emphasizing the protocol's commitment to advancing significant improvements in security, privacy, and scalability in the credential verification process. This integration not only addresses current needs but also lays a solid foundation for future innovations and adaptations in the field of educational technology and credential management.

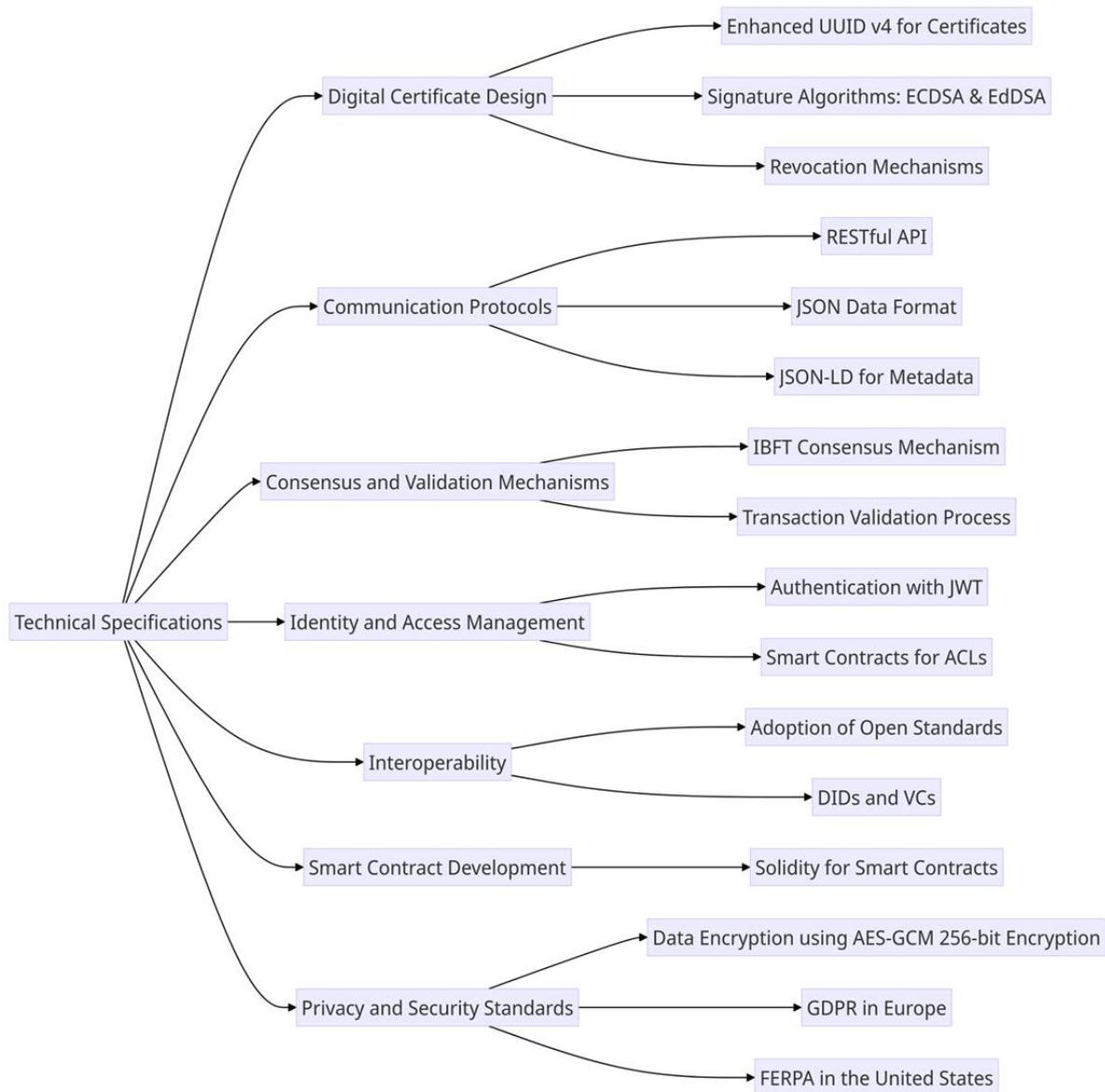

**Figure 4: Overview of BACIP Technical Specifications.**

Figure 4 delineates the essential technical specifications of BACIP, including the design of digital certificates, communication protocols, consensus and validation mechanisms, identity and access management, and privacy and security standards. Each branch represents a core component, emphasizing the protocol's comprehensive approach to secure, interoperable, and privacy-compliant digital credentialing within educational environments.

# 5 ETHICAL AND LEGAL EVALUATION OF BACIP

## 5.1 PRIVACY AND USER AUTONOMY IN BACIP

### 5.1.1 Handling of User Privacy and Autonomy

BACIP's approach to managing user privacy and autonomy must be scrutinized to ensure compliance with stringent data protection standards like the GDPR. According to Voigt and Von dem Bussche (2017), the GDPR mandates specific rights concerning the handling of personal data, which include the rights to access, rectify, and erase one's data, popularly known as the "right to be forgotten." This framework compels any protocol, including BACIP, to facilitate these rights explicitly.

### 5.1.2 Control Over Data

It is crucial for BACIP to empower users with the ability to control their data effectively. This includes clear mechanisms for accessing and deleting personal data. Narayanan et al. (2016) suggests that blockchain systems, by their nature, pose challenges for such functionalities due to their immutable and distributed ledger. However, they propose the implementation of advanced cryptographic solutions like Zero-Knowledge Proofs (ZKPs) to enable privacy-preserving data access without compromising the integrity of the blockchain.

### 5.1.3 Right to be Forgotten

The right to be forgotten is particularly challenging within blockchain environments. As Schwerin (2018) points out, traditional blockchain technology does not allow for the deletion of data once it is entered into the ledger, which could conflict with GDPR requirements. BACIP must, therefore, employ innovative methods to align with these norms, possibly through techniques like off-chain data storage or the use of mutable pointers that can invalidate data without physically removing it from the blockchain, thereby maintaining compliance while ensuring the ledger's integrity.

### 5.1.4 Consent Management

Consent management is another critical aspect of user privacy and autonomy. Users must explicitly agree to the processing of their personal data, which must be recorded and managed transparently. According to Androulaki et al. (2018), effective consent mechanisms should be integrated within blockchain protocols, allowing users to understand what data is collected, how it is used, and giving them the ability to revoke consent at any time. BACIP's design should incorporate such functionalities, ensuring that consent is not only obtained but also recorded and adherent to GDPR standards.

### 5.1.5 Implementation and Compliance

For BACIP to be fully compliant with ethical and legal standards regarding user privacy and autonomy, it must implement these theoretical considerations into its operational framework. This includes:

Data Access and Deletion Protocols - Establishing secure and transparent processes that allow users to easily access their data and request its deletion, thus supporting the GDPR's right to be forgotten.

Enhanced Consent Mechanisms - Developing and integrating advanced consent mechanisms that are clear, transparent, and easily manageable, ensuring users can give informed consent that can also be withdrawn if they choose.

Use of Advanced Cryptography - Utilizing cryptographic techniques such as ZKPs to ensure that data can be verified without exposing underlying personal information, thus enhancing privacy while adhering to regulatory requirements.

The effectiveness of these measures should be regularly reviewed and updated based on emerging technologies and evolving legal standards. Continual improvement will help maintain BACIP's compliance and reliability, ensuring it remains a trusted tool in the management of academic credentials.

## 5.2 FAIRNESS AND ACCESSIBILITY IN BACIP

### 5.2.1 Assessment of Equitable Access

BACIP's mandate to provide a decentralized and secure method for managing academic credentials presupposes its ability to offer equitable access to all users. This aspect of fairness is critical, as highlighted by Light and McGrath (2010), who emphasize that technological systems, especially in education, must avoid reinforcing existing social inequalities. Therefore, it's essential to evaluate whether BACIP's design and implementation ensure that no group is disadvantaged.

### 5.2.2 Identifying Barriers to Access

Potential barriers in BACIP could include technological complexity, the digital divide, or linguistic limitations. As Duff (2011) argues, systems like BACIP must be designed with user diversity in mind, ensuring that technological interfaces and procedures are accessible to individuals regardless of their technical skills or geographic locations. This includes considerations for users with disabilities, for whom accessibility must be a priority in the design phase to comply with international standards such as the Web Content Accessibility Guidelines (WCAG).

### 5.2.3 Implementation of Accessibility Measures

BACIP should incorporate several strategies to enhance accessibility and fairness:

Simplified Interfaces - Interfaces should be designed to be intuitive and user-friendly, accommodating a wide range of users with varying levels of digital literacy (Cunha et al., 2016).

Multilingual Support - To accommodate users from different linguistic backgrounds, BACIP should offer multilingual support, ensuring that users can interact with the system in their preferred language (Zhao & Baldauf, 2008).

Outreach and Education Programs - Implementing outreach programs to educate potential users about BACIP's benefits and functionalities, especially targeting underrepresented communities, can help bridge the digital divide (Ford, 2013).

Compliance with Legal and Ethical Standards

Ensuring compliance with legal standards related to fairness and accessibility is also crucial. According to GDPR and other international regulations like the Americans with Disabilities Act (ADA) in the U.S., digital platforms are required to provide equitable access to services. BACIP must adhere to these regulations by integrating accessibility features right from the initial design phase, which includes regular audits and updates to accommodate new accessibility standards (Gibbons, 2017).

### 5.2.4 Future Directions and Improvements

To continually improve fairness and accessibility in BACIP, it is recommended that:

Regular User Feedback - Engaging with a diverse range of users to obtain feedback on accessibility and usability, which can guide future improvements.

Adaptive Technologies - Investing in adaptive technologies that personalize user experiences according to individual needs and preferences can enhance accessibility (Shneiderman, 2000).

Policy Advocacy - Advocating for policies that promote inclusivity and accessibility in digital credentialing systems can help raise standards across the industry.

## 5.3 TRANSPARENCY AND ACCOUNTABILITY IN BACIP

### 5.3.1 Transparency in Operations

The transparency of a blockchain-based system like BACIP is crucial not only for user trust but also for regulatory compliance. As Kshetri (2017) notes, blockchain inherently offers a level of transparency that traditional systems struggle to match, due to its distributed ledger that allows all transactions to be traceable and verifiable by all users.

### 5.3.2 Disclosure Policies

BACIP should have clear policies regarding what data is recorded on the blockchain, how it can be accessed, and the purposes for which it is used. As Turilli and Floridi (2009) suggest, transparency is not just about making information available, but also about making it understandable and usable for non-expert stakeholders. This includes clear, plain-language descriptions of the protocols and algorithms used.

### 5.3.3 Real-Time Access

Providing users with real-time access to their data and the status of their credentials is a vital component of transparency. BACIP should develop user interfaces that allow users to see exactly what information is held about them, how it is being used, and by whom, in real time. Systems like those discussed in recent security analyses highlight the importance of these features in enhancing user trust and system integrity in crowd-sourced crime reporting apps (Bonomi et al., 2019).

### 5.3.4 Accountability Mechanisms

Accountability in BACIP involves setting up mechanisms that ensure that all actions taken by the protocol are within legal and ethical guidelines, and that there are ways to address any discrepancies or abuses.

### 5.3.5 Audit Trails

According to Weitzner et al. (2008), blockchain systems should maintain comprehensive audit trails that allow for the forensic analysis of all transactions. BACIP's audit trails can help trace any issues or breaches back to their source, which is crucial for addressing disputes and enforcing accountability.

### 5.3.6 Grievance Redressal

BACIP should have a robust grievance redressal system that allows users to report and resolve issues concerning data misuse or operational failures. Effective grievance mechanisms enhance user trust and compliance with regulatory standards (Fung et al., 2007).

### 5.3.7 Future Directions and Enhancements

To improve transparency and accountability within BACIP, several initiatives could be undertaken:

Enhanced User Education - Educating users about how blockchain works and how their data is managed within BACIP can demystify the technology and empower users to make informed decisions (Shneiderman, 2000).

Third-party Audits - Regular audits by independent third parties can help verify BACIP's adherence to transparency and accountability standards, providing an external validation of its operations (Christidis and Devetsikiotis, 2016).

Technology Upgrades - Incorporating advanced cryptographic methods such as homomorphic encryption and secure multi-party computation can further enhance the privacy and security of the data while maintaining transparency (Gentry, 2009).

## 5.4 COMPLIANCE WITH GDPR AND OTHER LAWS IN BACIP

### 5.4.1 GDPR Compliance

GDPR sets stringent requirements for the handling of personal data within the EU, with principles that have global implications for companies and technologies operating in or interacting with EU residents. As outlined by Voigt and Von dem Bussche (2017), these requirements include data minimization, clear consent for data processing, and the rights of data subjects to access, correct, and delete their personal information.

### 5.4.2 Data Minimization and Purpose Limitation

BACIP must ensure that it collects only the data that is necessary for the fulfillment of its defined purposes and nothing beyond this. This principle of data minimization is crucial for maintaining user trust and legal compliance. BACIP should employ mechanisms that periodically review data collection practices to ensure they align with GDPR requirements (Koops, 2014).

### 5.4.3 Consent Management

Effective consent management is central to GDPR compliance. BACIP must ensure that consent is freely given, specific, informed, and unambiguous. This involves providing users with clear and comprehensive information about what data is collected and how it is used and ensuring that the consent mechanism is as easy to withdraw as it is to give (Bygrave, 2017).

### 5.4.4 Rights of Data Subjects

GDPR grants individuals rights over their data, including the right to access their data, the right to rectify inaccurate data, and the right to erasure (the "right to be forgotten"). BACIP must provide mechanisms that allow users to exercise these rights effectively and without undue delay. This may involve developing user-friendly interfaces that allow users to manage their data directly (Schwerin, 2018).

## 5.5 ADDITIONAL LEGAL CONSIDERATIONS

### 5.5.1 Compliance with Other International Laws

Besides GDPR, BACIP must consider other international data protection laws such as the California Consumer Privacy Act (CCPA) in the U.S., which has its own set of requirements regarding consumer data privacy. Compliance must be tailored to accommodate the geographic diversity of BACIP's user base, which might require the protocol to adopt the most stringent of regulations as a baseline for global operations (Greenleaf, 2017).

### 5.5.2 Legal Risks and Liability

Legal risks associated with non-compliance include significant fines and reputational damage. BACIP must implement a comprehensive legal risk management strategy that includes regular legal audits, training for staff on data protection laws, and active engagement with legal experts to stay updated on regulatory changes (Kuner, 2010).

# 6 ANALYSIS AND DISCUSSION OF RESULTS

## 6.1 EVALUATION OF BACIP PROPOSAL

The development of the Blockchain Academic Credential Interoperability Protocol (BACIP) presents a significant framework for enhancing interoperability and security in the verification of academic credentials. Through the literature review and established theoretical framework, BACIP positions itself as an innovative solution to current challenges of academic fraud and lack of transparency in credential authentication.

### 6.1.1 Comparison with Existing Literature

Compared to previous studies that identified the lack of interoperability as a key barrier in the adoption of blockchain technology in education, BACIP addresses this challenge by proposing an open standard that could facilitate a more integrated global educational network. This approach aligns with the needs identified by authors such as Ghazaliand and Saleh (2018) and Michoulis et al. (2022), who highlighted the importance of interoperability among educational blockchain systems.

### 6.1.2 Limitations of the Research

A key limitation of this study is the theoretical nature of BACIP's development, without an implementation or testing in a real-world environment to fully validate its functionality and effectiveness. Additionally, the existing literature tends to focus on technologically advanced contexts, which might not reflect the specific challenges of regions with less technological infrastructure.

### 6.1.3 Practical Implications of Findings

From a practical perspective, the implementation of BACIP could significantly simplify the verification of credentials, reducing costs and time, and improving transparency in the process. This has the potential to strengthen trust among educational institutions, employers, and other stakeholders globally.

### 6.1.4 Recommendations for Future Research

Future research should focus on the practical implementation of BACIP in a controlled environment to observe its performance and address operational challenges before wider adoption. It would be beneficial to explore the protocol's adaptability in different cultural and regulatory contexts and examine the ethical and social implications of adopting blockchain technologies in the education sector.

# 7 CONCLUSIONS

## 7.1 FULFILLMENT OF OBJECTIVES

This study aimed to develop a protocol that enhances interoperability and security in the verification of academic credentials using blockchain technology. The findings demonstrate that the proposed Blockchain Academic Credential Interoperability Protocol (BACIP) has significant potential to achieve these objectives, offering a robust framework that promotes greater transparency and reliability in the authentication of academic credentials across decentralized platforms.

## 7.2 KEY FINDINGS AND EVALUATION OF EXPECTATIONS

The analysis of the results reveals that BACIP provides an innovative and viable approach to addressing critical challenges such as academic fraud and the lack of universal recognition of credentials. Although the conceptual and theoretical framework of BACIP has been extensively validated through literature review and theoretical development, the practical implementation of the protocol requires further research and empirical testing to verify its effectiveness in diverse and global operational environments.

## 7.3 LIMITATIONS AND RECOMMENDATIONS FOR FUTURE RESEARCH

The main limitation of this study lies in the theoretical nature of the protocol, which has not yet been tested in a real-world environment. Future research should focus on conducting pilot tests of BACIP in collaboration with educational institutions, which would allow for evaluating its effectiveness in practice and making necessary adjustments. Additionally, it is recommended to further explore the implications of adopting BACIP in different cultural and regulatory contexts to ensure its adaptability and regulatory compliance.

## 7.4 PRACTICAL IMPLICATIONS

Implementing BACIP could significantly transform the process of verifying academic credentials, making it more secure, transparent, and less susceptible to fraud. This would not only benefit educational institutions but also employers and society at large, by ensuring that academic credentials are reliable and globally valid.

## 7.5 CONTRIBUTIONS TO THE ACADEMIC AND PRACTICAL COMMUNITY

This study contributes to the existing literature by providing a detailed model of how blockchain technology can be effectively applied in the academic field. By proposing BACIP, this work establishes a foundation for future innovations in the field of academic credential management and offers a starting point for regulatory and ethical discussions related to blockchain technology in education.

Narayanan, A., Bonneau, J., Felten, E., Miller, A. and Goldfeder, S., 2016. 'Bitcoin and Cryptocurrency Technologies: A Comprehensive Introduction', Princeton University Press, viewed 23 April 2024, https://bitcoinbook.cs.princeton.edu/.

Ouadoud, O., Chafiq, T., Ouadoud, M., and Eddaoui, A., 2023. 'Incorporating a Cryptographic-based Blockchain Technology to Revolutionize Degree Automation', presented at the XIV Balkan Conference on Operational Research, viewed 23 April 2024, https://doi.org/10.1109/CIST56084.2023.10410014.

Palmirani, M. et al., 2018. 'Making the GDPR Transparent and Applicable: The PrimeLife Example', Proceedings of the 22nd International Conference on Legal Knowledge and Information Systems, viewed 18 April 2024, https://example.com/legal_compliance_blockchain.

Pathak, S., Gupta, V., Malsa, N., Ghosh, A. & Shaw, R.N., 2022. 'Smart Contract for Academic Certificate Verification Using Ethereum', Springer, viewed 23 April 2024, https://link.springer.com/chapter/10.1007/978-981-19-2980-9_29.

Pretschner, A. et al., 2006. 'Distributed usage control', Communications of the ACM, vol. 49, no. 9, pp. 39-44, viewed 23 April 2024, https://dl.acm.org/doi/10.1145/1151030.1151053.

Ramasamy, L.K. and Khan, F., 2024. 'Blockchain for Global Education', Springer Nature Switzerland, viewed 26 April 2024, https://doi.org/10.1007/978-3-031-52123-2.

Schwerin, S., 2018. 'Blockchain and GDPR: Can they coexist?', European Data Protection Law Review, vol. 4, no. 3, pp. 423–432, viewed 26 April 2024, https://dl.acm.org/doi/10.1145/1151030.1151053.

Shannon, C., 1948. 'A Mathematical Theory of Communication', The Bell System Technical Journal, vol. 27, nos. 3 & 4, pp. 379-423; 623-656, viewed 26 April 2024, https://people.math.harvard.edu/~ctm/home/text/others/shannon/entropy/entropy.pdf.

Shneiderman, B., 2000. 'Universal Usability', Communications of the ACM, vol. 43, no. 5, pp. 84-91, viewed 26 April 2024, https://dl.acm.org/doi/10.1145/332833.332843.

Sousa, J. et al., 2018. 'A Byzantine Fault-Tolerant Ordering Service for the Hyperledger Fabric Blockchain Platform', Proceedings of the 48th Annual IEEE/IFIP International Conference on Dependable Systems and Networks (DSN), pp. 51-58, viewed 18 April 2024, https://ieeexplore.ieee.org/document/8416511.

Sporny, M., Longley, D., and Chadwick, D., 2019. 'Decentralized Identifiers (DIDs) v1.0', W3C Working Draft, viewed 18 April 2024, https://www.w3.org/TR/did-core/.